\documentstyle[aps,prl]{revtex}

\input{psfig}

\begin{document}

\twocolumn[\hsize\textwidth\columnwidth\hsize\csname@twocolumnfalse\endcsname

\title{Pd/Cu Site Interchange and Non-Fermi-Liquid Behavior in UCu$_4$Pd}

\author{C. H. Booth,$^{1,}$\cite{Booth_email} D. E. MacLaughlin,$^{1,2}$ 
R. H. Heffner,$^1$ R. Chau,$^{3,}$\cite{LLNL} M. B Maple,$^3$ and 
G. H. Kwei,$^1$} 

\address{$^1$Los Alamos National Laboratory, Los Alamos, New Mexico 87545}
\address{$^2$Department of Physics, University of California,
Riverside, California 92521}
\address{$^3$Department of Physics, University of California, San Diego, 
California 92093}

\date{to be published in Phys. Rev. Lett., LA-UR-98-3125}

\maketitle

\begin{abstract}

X-ray-absorption fine-structure measurements of the local structure in
UCu$_4$Pd are described which indicate a probable lattice-disorder origin
for non-Fermi-liquid behavior in this material. Short Pd-Cu distances are
observed, consistent with $24 \pm 3$\% of the Pd atoms occupying nominally
Cu sites. A ``Kondo disorder'' model, based on the effect on the local
Kondo temperature~$T_K$ of this interchange and some additional bond-length
disorder, agrees quantitatively with previous experimental susceptibility data, 
and therefore also with specific heat and magnetic resonance experiments.

\end{abstract}



]

\narrowtext


A number of heavy-fermion $f$-ion compounds and alloys exhibit
low-temperature thermodynamic and transport properties which do not behave
as expected from the Landau Fermi-liquid theory thought to be applicable to
heavy-fermion systems \cite{NFL_mat_ex}. Typical non-Fermi liquid (NFL) anomalies
at low temperatures include logarithmic or weak power law dependences of the
Sommerfeld specific heat coefficient~$\gamma(T) = C(T)/T$ and the magnetic
susceptibility~$\chi(T)$, and a linear temperature dependence of the
electrical resistivity
$\Delta \rho(T) = \rho(T) - \rho(0)$ rather than the $T^2$ dependence expected 
from Fermi-liquid
theory. Most but not all NFL materials are disordered alloys, and
theoretical treatments of NFL behavior have differed as to 
whether this disorder is important. 

Recent nuclear magnetic resonance (NMR)\cite{Bernal95} and muon-spin rotation 
($\mu$SR)\cite{MacLaughlin96} experiments revealed a broad inhomogeneous 
distribution of the
low-temperature local susceptibility in UCu$_{5-x}$Pd$_x$, $x = 1.0$ and
1.5. A phenomenological ``Kondo disorder'' model 
can account for the NFL behavior as due to a broad
distribution~$P(T_K)$ of Kondo temperatures~$T_K$, such that a significant
number of $f$-ion spins with very low $T_K$ remain uncompensated at low
temperatures. These uncompensated spins dominate the low-temperature
thermodynamic properties and are able to account for the $T$-linear
resistivity \cite{Miranda97}. This model 
also 
describes the specific heat and susceptibility of the disordered NFL
alloy~CeRh$_{2-x}$Ru$_x$Si$_2$, $x = 1.0$ \cite{Graf97}. Despite the ability
to account for many of the electronic properties of these materials, 
evidence for the microscopic origin of Kondo disorder is still
necessary to confirm that it properly describes the physics of these materials.

The present study aims to determine the extent to which structural
disorder may cause Kondo disorder in a nominally ordered NFL system. 
We chose UCu$_4$Pd for this purpose, since it is well 
studied \cite{Bernal95,MacLaughlin96} and has recently been 
characterized by neutron diffraction \cite{Chau98}.  In this diffraction study, 
Pd is shown to have a strong tendency to occupy the $4c$ site in the 
cubic $\bar{4}3m$ structure.  However, the experiment cannot differentiate 
between full $4c$ site occupancy by Pd and up to approximately 16\% of the $4c$ 
sites occupied by Cu rather than Pd atoms. Using a simple
heuristic model, we estimate below that Pd/Cu site interchange of 
this magnitude may be enough to cause sufficient Kondo disorder to
explain the NFL behavior.  A better measurement of
the amount of site interchange is difficult for neutron measurements, since the 
contrast between the  coherent scattering lengths for U (8.4 fm), Pd (5.9 fm), 
and Cu (7.7 fm) is not large. 

A better way to determine the extent of any Cu/Pd site interchange in UCu$_4$Pd 
is to use a local probe to look at the near-neighbor bond-length distributions. 
If Pd atoms sit on 
some Cu sites, for instance, there will be Pd-Cu distances at $\sim 2.50$ \AA\ 
overlapping Cu-Cu pairs at the same distance.  It is therefore vitally important
to use a probe that is atomic-species specific.  For this reason we have 
performed x-ray-absorption fine-structure (XAFS) experiments on the same sample 
of UCu$_4$Pd used in Ref. \cite{Chau98}, using the U $L_{\rm III}$ edge 
and the Pd and Cu $K$ edges.


Details of the sample-growth procedure are given in Ref. \cite{Chau98}.  
After the $\mu$SR experiments \cite{MacLaughlin96}, approximately 200~mg of a 
pellet was soaked in methanol to 
remove a binding varnish, and reground by hand in acetone. The average particle 
size was estimated from microscopy to be $\sim 20 \mu$m.  In addition,
the sample was passed through a 20 $\mu$m sieve for the Cu-edge experiment.
The sample was 
then brushed onto scotch tape, and layers of the tape were stacked 
to 
produce a change in the absorption coefficient of approximately unity.  

X-ray absorption data were collected at the Stanford Synchrotron Radiation 
Laboratory from the U $L_{\rm III}$ and Pd $K$ edges at BL 10-2 using Si(220) detuned 
double monochromator crystals.  The Cu $K$-edge data were collected at BL 2-3 
with Si(111) crystals, similarly detuned. The vertical slit height was set at 
0.7 mm.  Data were collected at $T$=3.3~K for the U and Pd edges, and at 
$T$= 20~K for the Cu edge, with the temperature controlled within 0.1~K\@.  Two 
scans each were taken for the U and Pd edges, while three scans were
collected for the Cu edge data.  More scans were collected at temperatures 
ranging up to 320~K; these will be discussed in a future article. The absorption
data were reduced using standard procedures \cite{Li95b}, which include fitting 
an embedded-atom absorption $\mu_0(E)$ with a 5 knot spline function.  The XAFS 
function $\chi(k)$ is defined as $\mu(k)/\mu_0(k)-–1$, where $\mu$ is the 
absorption coefficient, $k=\sqrt{2m_e/\hbar^2}(E-E_0)$ is the photoelectron 
wave vector, $m_e$ is the electron rest mass,  and the 
ionization threshold energy $E_0$ is chosen arbitrarily as the energy at the 
half-height of the absorbing edge, and is allowed to vary in subsequent fits.

\begin{figure}[t]
\psfig{file=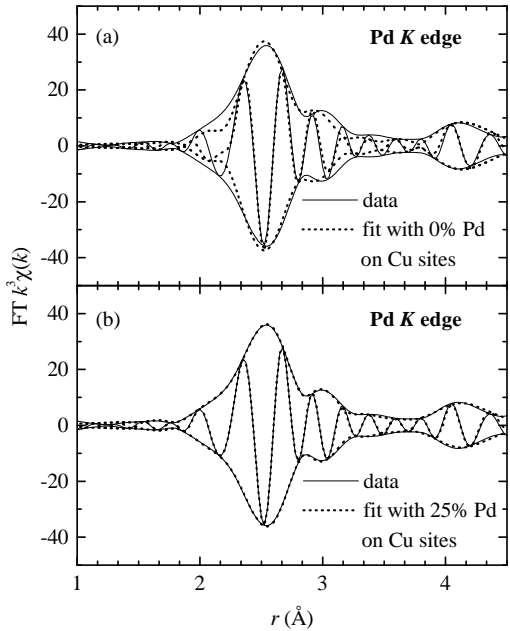,width=3.0in,clip=}
\vspace{-2ex}
\caption{Fourier transform (FT) of $k^3\chi(k)$ from the Pd $K$ edge.  The
outer envelope is the amplitude and the oscillating inner line is the real part 
of the complex transform.  Solid lines are data taken at $T$=3.3~K, and the 
dotted lines are fits assuming (a) only 
bonds found in the nominally ordered structure, and (b) also bonds consistent
with Pd on Cu sites.  Transforms are from 
3.0-14.5 \AA$^{-1}$, and Gaussian broadened by 0.3 \AA$^{-1}$.  Fits are from
2.0 to 4.5 \AA.
}
\label{Pd_rs_fig}
\end{figure}


XAFS provides information about the atomic arrangement of near neighbor atoms 
to an absorbing atom.  The XAFS function $\chi(k)$ has oscillations 
due to the interference of the outgoing photoelectron and the component 
back scattered to the absorbing atom.  By fitting with theoretical 
calculations of the back-scattering amplitude and phase for each 
shell \cite{FEFF6} we can determine the parameters $S_0^2$, $N_i$, $R_i$, and 
$\sigma$, defined as follows.
$S_0^2$ is an amplitude reduction factor, which is poorly determined by theory 
and is therefore fit somewhat arbitrarily.  It is correlated in the fits
with the 
number of neighbors in the $i$th shell $N_i$, so we hold $N_i$ fixed whenever 
possible. The bond length to the $i$th shell $R_i$ and the distribution width
$\sigma_i$ of this shell are better determined.  More details of fitting 
procedures are found in 
Ref. \cite{Li95b}.

Before describing the data, we first recall that the crystal structure of 
UCu$_4$Pd can be described as interlocking U and Pd fcc lattices, 
with an interstitial vertex-sharing network of Cu tetrahedra.  From a local 
perspective, the U and Pd sites are very similar.  Each site has twelve Cu 
neighbors at 2.93 \AA\ made up from the faces of four Cu tetrahedra, and 
four Pd (for U sites) or U (for Pd sites) neighbors at 3.06 \AA.  The local
Cu environment is made up of four Cu-Cu pairs at $\sim 2.50$ \AA\ and four
each Cu-U and Cu-Pd pairs at $\sim 2.93$ \AA.

Since the Pd $K$-edge data determine the model we will use for the 
distortion, we start with a discussion of these data and then present the U 
and Cu edge results.  First consider a fit to the data that does not 
include any site interchange.  We assume that the $N_i$ are given by the ordered
structure, so that the overall amplitude is given only by $S_0^2$ and the 
widths $\sigma_i$ for the individual shells.  Such a fit is displayed in 
Fig.~\ref{Pd_rs_fig}(a).  The main peak in this transform at $\sim 2.6$ \AA\ is
primarily due to the Pd-Cu pairs, which are separated by 2.93 \AA.  (Peaks in
XAFS transforms are shifted from the atom-pair distances by a 
phase shift of the photoelectron at both the absorbing and the back-scattering 
atoms). The shoulder on the 
right side is from the Pd-U path at 3.06 \AA.  This fit is of very poor quality,
particularly on the low side of the main Pd-Cu peak.  If we then allow for a 
model with some Pd on the Cu sites (the main effect arises from adding a Pd-Cu bond 
at $\sim 2.50$ \AA), we are much better able to fit the low side of the main 
peak.  In fact, the fit [Fig.~\ref{Pd_rs_fig}(b)] converges with 
bond lengths for the extra added 
peaks which are very close to the bond lengths from the copper site. Using this 
fit as a reference, the undistorted fit has a worse statistical-$\chi^2$ by over
3000\%!  Moreover, since each back-scattering atomic species has a distinctive 
phase and back-scattering amplitude, we know that the extra peaks indeed involve
the species 
consistent with Pd/Cu site interchange.  We estimate the fraction of Pd atoms
that reside on Cu sites to be $24\pm3$\%.

Fits to the U and Cu edge data are more ambiguous about the 
Pd/Cu site interchange observed from the Pd edge.  The
signal from the more abundant Cu atoms and overlapping shells make
fits that include Pd/Cu site interchange only slightly better than those that 
do not.  U/Cu site interchange, on the other hand, should be detectable from the
U edge data by observing a short U-Cu peak.  We measure no such peak,
and place a limit of $\leq$ 10\% of the U atoms on Cu sites.
All these fits are consistent with the average structure determined by 
diffraction, and are the same quality as the fit in Fig.~\ref{Pd_rs_fig}(b).


Based on the measurements presented above, we now know that a significant
fraction of the Pd atoms ($\sim 1/4$) sit on the nominally Cu $16e$ sites,
indicating that this system is not intrinsically homogeneous.  This result 
therefore strongly suggests the applicability of Kondo disorder models  
for UCu$_4$Pd,
and that the microscopic origin of this disorder
is structural.  With this knowledge, we are compelled to ask, is the
measured amount of disorder consistent with the observed NFL behavior?

\begin{figure}[t]
\psfig{file=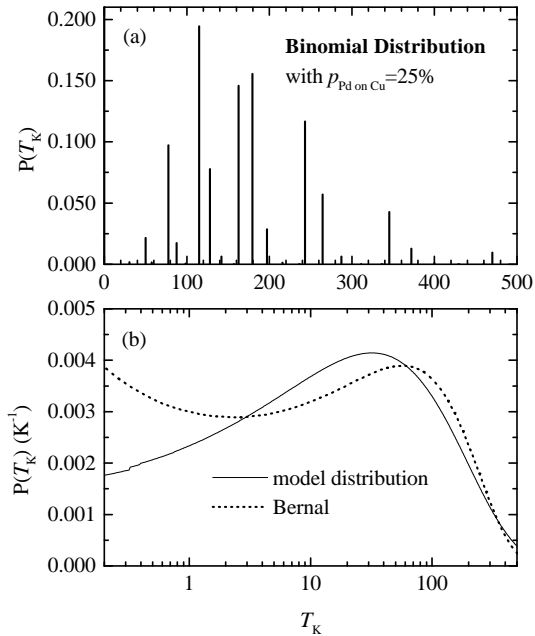,width=3.0in,clip=}
\caption{(a) Distribution of $T_K$'s based on 25\% probability of a given Pd on 
a Cu site, as described in the text.  (b) Distribution for (a) with 0.047 \AA\ 
additional bond length disorder. The
distribution from Ref. \protect\cite{Bernal95} is shown for
comparison.
}
\label{distrib_fig}
\end{figure}

To answer this question, we calculate a distribution of Kondo temperatures 
$P(T_K)$, and from this distribution and a Bethe-ansatz calculation of the 
susceptibility of single Kondo impurity, we calculate the susceptibility
$\chi(T,P(T_K))$.  We apply disorder in a local fashion to the
Kondo temperature:
\begin{equation}
T_K=E_F e^{-1/(\rho {\cal J})},
\label{TK_eqn}
\end{equation}
where $E_F$ is the Fermi energy, $\rho$ is the density of 
states at the Fermi level and $\cal J$ is the conduction-electron/local-moment 
exchange energy.  
We hypothesize that $\rho$ is primarily a 
long-range quantity, and consider only disorder in $\cal J$.  To this end, 
we employ a tight-binding 
formula\cite{Harrison87} for the coupling energy between the electronic shells
of neighboring atoms.
The dominant contribution is from $f{\rm-}d$ hybridization, given by:
\begin{equation}
V_{fd}=\frac{\eta_{fd} \hbar^2}{m_e}\frac{(r_{{\rm U}f}^5 
r_{{\rm X}d}^3)^{1/2}}{R_{\rm U-X}^6},
\label{V_eqn}
\end{equation}
where $r_{X\ell}$ is the radius of the electronic shell with angular momentum 
$\ell$ for 
atom $X$ (tabulated in Ref. \cite{Straub85}), and $R$ is the bond length between
the atoms with $\ell =f$ and $d$'.  The coefficient $\eta_{fd}$ depends only on 
the $\ell$'s and the bond symmetry (see Appendix B in Ref. \cite{Harrison87}). 
We will assume all bonds are
$\sigma$-bonds.  This coupling energy is related to the exchange energy
in Kondo theory simply by ${\cal J} = V_{fd}^2/\epsilon_f$, where 
$\epsilon_f$ is the $f$-level energy.  Within this tight-binding approach,
calculated energies are often too high compared to experiment by a factor
of two \cite{Harrison87}.  

As our basic model of the lattice, we sum Eq. (\ref{V_eqn}) over U-$X$ pairs 
that have a significant hybridization $V_{fd}$.  These include the twelve 
nearest neighbor U-Cu pairs at 2.93 \AA, and the four U-Pd pairs at 3.06 \AA.  
In fact, the hybridization per bond is higher for the U-Pd pairs because the 
radius of the $d$ shell of Pd is nearly 50\% larger than that of Cu.  To this 
base structure, we then allow each Pd site to have a 25\% chance of having a Cu 
atom on it, and each Cu site to have a 6.25\% (25/4) chance of having a Pd on 
it.  The effect of Pd on Cu sites is to raise the local hybridization, and thus 
the local $T_K$.  Replacing Pd with Cu on Pd sites has the opposite effect.  All
the different possible configurations and their probabilities (as given by a 
binomial distribution) were calculated.  The resulting discrete distribution of 
$T_K$'s is shown in Fig.~\ref{distrib_fig}(a).  We use $E_F$ from 
Ref. \cite{Bernal95} and $\rho/\epsilon_f=$7.27~eV$^{-2}$, as described below.
Notice that there is no mechanism within 
this model to provide any weight within the binomial model at $T_K$ 
near 0~K; Fig.~\ref{distrib_fig}(a) reflects this inadequacy.

In order for this model to be complete, however, we must also consider the
effect of bond length disorder on the hybridization, as occurs from normal 
(Debye) vibrations and possibly other static or dynamic origins.  We 
estimate this disorder
for an arbitrary distribution of $V_{fd}$ by calculating an average bond length
$R^\prime$ (assuming $r_d$ for Cu) and convolving Eq. (\ref{V_eqn}) with a 
distribution of bond lengths with half-width $\sigma$.  We write the 
pair-distribution function as  $N_\phi e^{-\phi(r)/(k_BT)}$, where $N_\phi$ is a 
normalization factor and $\phi(r)$ is a suitable interatomic potential.
We choose a Lennard-Jones potential rather than a harmonic or a Morse potential
because it models the hard-core repulsion and still only has two input 
parameters.  We define the position of the peak in the distribution as 
$R^\prime$ and the inverse of the  coefficient of the second-order term 
as $\sigma^2$.
Convolving this distribution with any $V_{fd}$, one finds the spread in $T_K$'s 
to be surprisingly large.  To illustrate the magnitude of this effect, 
consider the case of UCu$_4$Pd without site interchange (contrary to
our measurement), but with a modest amount of static disorder.
We estimate as little as $\sigma=0.01$ \AA\ of static disorder
generates a spread of Kondo temperatures 
$\sigma_{T_K}/\langle T_K \rangle \sim 40$\%.  

We calculate $\chi(T,P(T_K))$ by convolving the 
present $P(T_K)$ using 25\% site interchange and bond length disorder $\sigma$
with a theoretical $\chi(T,T_K)$ using the 
Bethe ansatz\cite{Rajan83}.  This curve is for a $J$=3/2
state with a Land\'{e} $g$-factor of 1.8745, and we hold $E_F=1.41$~eV, as used 
in Ref. \cite{Bernal95}.  This procedure requires $\rho/\epsilon_f$ and
$\sigma$ as fitting parameters.
We fit this distribution to the susceptibility derived from the $P(T_K)$
in Ref. \cite{Bernal95}, since it is continuous and essentially fits the data 
exactly down to $\sim 2$~K\@. 
The results for the final fit parameters are 
$\rho/\epsilon_f=7.27\pm0.07$~eV$^{-2}$ and $\sigma=0.047\pm0.002$ \AA.
The latter parameter is within the range of the static 
distortion estimated from the U $L_{\rm III}$-edge data.  Indeed, some level of 
static disorder is expected for a system with such a high degree of site 
interchange.  However, the amount of Pd/Cu site 
interchange and $\sigma$ are very strongly correlated such that good fits are
obtainable even with no site interchange and a slightly larger 
($\sim 0.002$ \AA)
value of $\sigma$.  Therefore, we cannot use the susceptibility data
alone to determine the amount of site interchange.

\begin{figure}[t]
\psfig{file=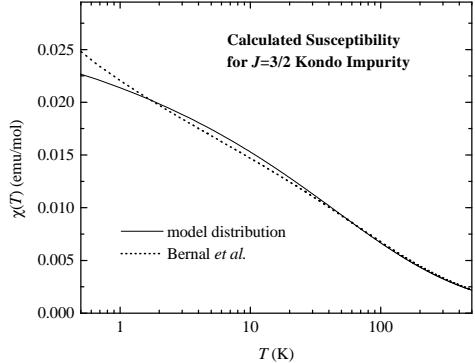,width=3.0in,clip=}
\caption{Magnetic susceptibility $\chi(T)$ calculated by convolving $P(T_K)$
from Fig.~\protect\ref{distrib_fig}(b) with a $J$=3/2 impurity from 
Ref. \protect\cite{Rajan83}.
}
\label{chi_fig}
\end{figure}

The final $P(T_K)$ distribution is shown in 
Fig.~\ref{distrib_fig}(b) and the susceptibility fits are shown in 
Fig.~\ref{chi_fig}.
The agreement between the susceptibility curves is very good for $T_K$ larger than 
$\sim 1$~K\@. Since the actual data only go down to $\sim 2$~K, the
distribution we are testing is consistent with these data.  In any case,
the discrepancy 
between the models may be difficult to interpret since 
it depends on the theoretical calculations for both the
susceptibility of a Kondo impurity and the $f{\rm -}d$ hybridization.

Despite these shortcomings, the present model has some good fundamental 
features that differ from the previous model used in Ref. \cite{Bernal95}.  One 
difference is that the present distribution of $\rho{\cal J}$ is chosen from a 
microscopic model based on experiment, rather than arbitrarily chosen to be 
Gaussian.  The choice of a Gaussian causes $P(T_K)$ in 
Ref.~\cite{Bernal95} to diverge as 
$T_K \rightarrow$ 0~K\@.  The distribution used in the present work cannot diverge
as $T_K \rightarrow0$~K because $V_{fd} \rightarrow 0$ only for 
$R \rightarrow \infty$.  
This lack of divergence should cause $\chi(T)$ as $T \rightarrow 0$~K to
reach a limiting value.  Such behavior has in fact been observed at temperatures
below $\sim 0.2$~K \cite{Vollmer97}, although we do not consider the alleged 
spin-glass behavior.  Finally, we must point out that more sophisticated models
of Kondo disorder, such as the recently proposed Griffiths 
phase \cite{Neto98} are possibly consistent with the work presented
here.

In summary, we have measured the local structure around U, Cu, and Pd atoms
using XAFS in UCu$_4$Pd.  The Pd data show a significant number of Cu bonds at a
distance consistent with 25\% of the Pd atoms on Cu ($16e$) sites.  
This level of lattice disorder strongly suggests the Kondo disorder
model as a starting place for understanding NFL behavior in this compound.  To
test this assertion, we present a model of the distribution of Kondo 
temperatures assuming a binomial distribution of Pd on Cu sites (and vice 
versa), together with some disorder in the mean bond length around each U atom. 
Hybridization strengths are calculated assuming a tight-binding model.  
The model is shown
to yield a susceptibility quantitatively similar to the measured $\chi(T)$ of 
Ref. \cite{Bernal95}.  Since the specific heat, NMR and $\mu$SR results can
also be described in terms of $P(T_K)$\cite{Bernal95,MacLaughlin96}, this 
model can describe many of the electronic properties.  This ability combined 
with the Cu/Pd site interchange measurements
is very strong evidence in favor of lattice disorder as the microscopic
mechanism of NFL behavior in UCu$_4$Pd.


The authors thank A. L. Cornelius, J. M. Lawrence, M. F. Hundley, and 
J. D. Thompson for useful conversations, and F. Bridges 
for assistance in collecting the data.  Work at Los Alamos was
conducted under the auspices of the U.S. Department of Energy (DOE). We 
acknowledge the support of the U.S. National Science Foundation,
Grants DMR-9418991 (D. E. M.) and DMR-9705454 (M. B. M., R. C.).
The experiments were performed at SSRL, which is operated by the
DOE, Division of Chemical Sciences, and by the NIH, Biomedical
Resource Technology Program, Division of Research Resources.

\end{document}